\documentclass[12pt]{article}
\usepackage{hyperref}

\usepackage[vcentermath]{youngtab}
\usepackage{color}

\def\be{\begin{equation}}
\def\ee{\end{equation}}
\def\ba{\begin{eqnarray}}
\def\ea{\end{eqnarray}}

\def\qr{{\rm q}}

\def\beq{\begin{equation}}
\def\eeq{\end{equation}}
\def\bea{\begin{eqnarray}}
\def\eea{\end{eqnarray}}

\def\m{\mu}

\usepackage[centertags]{amsmath}
\usepackage{amsfonts} 
\usepackage{amssymb}
\usepackage{amsthm}
\numberwithin{equation}{section} 
\def\ii{{\rm i}}

\begin{document}
\begin{titlepage}
\hfill \hbox{CERN-PH-TH-2015-018}
\vskip 0.1cm
\hfill \hbox{NORDITA-2015-13}
\vskip 0.1cm
\hfill \hbox{QMUL-PH-15-02}
\vskip 1.5cm
\begin{flushright}
\end{flushright}
\vskip 1.0cm
\begin{center}
{\Large \bf   Regge behavior saves String Theory \\ from causality violations}
 
  \vskip 1.0cm {\large Giuseppe
D'Appollonio$^{a}$, Paolo Di Vecchia$^{b, c}$,
Rodolfo Russo$^{d}$, \\
Gabriele Veneziano$^{e, f}$ } \\[0.7cm]
{\it $^a$ Dipartimento di Fisica, Universit\`a di Cagliari and
INFN, Sezione di Cagliari\\ Cittadella
Universitaria, 09042 Monserrato, Italy}\\
{\it $^b$ The Niels Bohr Institute, University of Copenhagen, Blegdamsvej 17, \\
DK-2100 Copenhagen, Denmark}\\
{\it $^c$ Nordita, KTH Royal Institute of Technology and Stockholm University, \\Roslagstullsbacken 23, SE-10691 Stockholm, Sweden}\\
{\it $^d$ Queen Mary University of London, Mile End Road, E1 4NS London, United Kingdom}\\
{\it $^e$ Coll\`ege de France, 11 place M. Berthelot, 75005 Paris, France}\\
{\it $^f$Theory Division, CERN, CH-1211 Geneva 23, Switzerland}
\end{center}
\begin{abstract}
Higher-derivative corrections to the Einstein-Hilbert action are present in  bosonic string theory  leading to the potential causality violations recently pointed out by Camanho et al.~\cite{Camanho:2014apa}.
We analyze in detail this question by considering high-energy string-brane collisions at impact parameters $b  \le l_s$ (the string-length parameter) with $l_s \gg R_p$ (the characteristic scale of the D$p$-brane geometry).
 If we keep only the contribution of the massless states causality is violated for a set of initial  states whose polarization is suitably chosen with respect to the impact parameter vector.  Such violations are instead neatly avoided when the full structure of string theory --and in particular its Regge behavior-- is taken into account.
\end{abstract}
\end{titlepage}

\tableofcontents

\section{Introduction }
\label{1}
\label{intro}

In a recent paper \cite{Camanho:2014apa} Camanho, Edelstein, Maldacena, and Zhiboedov (CEMZ) have considered higher-derivative corrections to the Einstein-Hilbert action of the type
\beq
S = \frac{1}{16 \pi G} \int d^dx \sqrt{-g} (R +l_2^2 R^2 + l_4^4 R^3) \; \ , 
\label{higherder}
\eeq
where the $R^2$ ($R^3$) corrections actually stand for precise four (six) derivative corrections that guarantee the absence of second-order time derivatives, and $l_2, l_4 $ denote new length scales  below which such corrections become non negligible. Moreover, they assumed that $l_2$ and $l_4$ are much larger than the Planck length $l_P = (G\hbar)^{1/(d-2)}$ to ensure that there is an intermediate  energy regime where the higher derivative corrections are important, but the theory is still weakly coupled.
 
 CEMZ  argued that, through their modification of the three-graviton vertex, these corrections can easily lead to
 short-distance violations of causality by inducing (Shapiro) time delays of the wrong sign for some carefully chosen gedanken experiments.  They go on to show that this problem cannot be fixed unless one assumes that the full theory contains  an infinite number of states extending to arbitrarily high spin. They finally use  results about high energy scattering in superstring theory \cite{Amati:1987uf} to illustrate this possible way out of the causality problem.
 
 The Gedanken experiment chosen by CEMZ is the high-energy scattering of a polarized ``probe" graviton off a coherent state ``target "  consisting of $N$-polarized gravitons. The regime is chosen so  that the full $S$ matrix becomes the $N$th power of an almost trivial elastic two-body $S$-matrix. In a suitable large-$N$ limit the  $S$-matrix exponentiates and  builds up an eikonal-like phase $2 \delta(E,b)$ (where $E$ is the probe-graviton energy and $b$ the impact parameter) from which it is easy to compute the time delay as
 \beq
 \label{tdelayC}
 \Delta t = 2 \partial_E \delta(E,b) = ( \Delta t)_{EH} \left( 1 \pm c_2 \frac{l_2^4}{b^4}   \pm c_4 \frac{l_4^8}{b^8}  \right) \ , 
 \eeq
 where $c_2, c_4$ are some numerical constants and the $\pm$ sign choices  depend on the relative orientation of the helicities of the probe and of the target. Because of these sign choices it is always possible to choose the relative polarizations in such a way that, at $b \ll l_2, l_4$, the time delay has the opposite sign of the usual (Einstein-Hilbert) time delay $ ( \Delta t)_{EH}$. This is, in a nutshell, the causality 
violation claimed in   \cite{Camanho:2014apa}.
 
 The string theory counterexample presented in \cite{Camanho:2014apa} is the high energy collision of two gravitons
 at small impact parameter and at sufficiently small string coupling for the Schwarzschild radius $R_S$ associated 
with the center of mass energy to be much smaller than the string-length parameter $l_s$.
As it is well known, $R^2$ and $R^3$ corrections are both absent in the case of the maximally supersymmetric string theory. The heterotic string 
has only $R^2$ corrections, while the bosonic string has both $R^2$ and $R^3$ corrections. For this reason in this paper we analyze in detail the Shapiro time delay in the bosonic string case. This is possible since the closed string tachyon present in the spectrum is harmless in the Regge limit as briefly discussed in Section~\ref{4}. We expect the analysis of string-string scattering in the heterotic case to be very similar.

In order to make the connection with CEMZ as close as possible, we will consider the case of a probe graviton scattering on a heavy target consisting, in our case, of a 
 stack of $N \gg1$ D$p$-branes.  In the superstring case this process has been studied recently in
  \cite{D'Appollonio:2010ae}, \cite{Black:2011ep}, \cite{D'Appollonio:2013hja} (see also \cite{D'Appollonio:2013rsa}) where the main aim was to recover classical phenomena such as gravitational deflection and tidal excitation in the large-$b$ regime.
  
Here, besides switching to the bosonic string case, we also look (again in order to make contact with CEMZ) at the $b \ll l_s$ regime and with small enough string-coupling (Newton constant) for the radius $R_p$ of the brane geometry to be smaller than $l_s$. This set up will allow us to illustrate in a simple and transparent way the emergence of the CEMZ causality problem in a properly defined field theory limit as well as its resolution in 
the full-fledged string theory.
The rest of the paper is organized as follows.

In Section 2 we derive, for the critical closed bosonic string,  the  tree-level elastic scattering amplitude of a massless string (a graviton, a dilaton or a Kalb-Ramond boson) impinging on a stack of D$p$-branes and analyze its Regge asymptotics.  We then argue that, in analogy with what was already shown for the superstring~\cite{D'Appollonio:2010ae}, \cite{D'Appollonio:2013hja}, open string loop corrections lead to a unitary S-matrix in the form of an operator eikonal expression.

In Section 3 we discuss an appropriate QFT truncation of the results described in Section 2 corresponding to a field theoretic gravity theory 
with higher derivative corrections containing, besides a graviton $G_{\mu\nu}$, an antisymmetric (Kalb-Ramond) field $B_{\mu\nu}$ and a dilaton $\phi$. We then show that such a theory suffers from the causality (and equivalence principle) violations pointed out in~\cite{Camanho:2014apa}.

In Section 4 we show in detail how the above mentioned problems are neatly avoided when the full structure of string theory (in particular its Regge behavior) is taken into account. Section 5 gives a summary of our findings as well as some short conclusions.

Appendix A contains a sketch of the calculation of the tree-level amplitudes presented in Section 2 while Appendix B
gives a check of the operator eikonal exponentiation at one-loop order (annulus diagram). In Appendix C we give, for the sake of completeness, the derivation of the Shapiro time delay for the processes discussed in this paper.


\section{Bosonic string amplitudes in presence of D-branes}
\label{2}
\setcounter{equation}{0} \label{strampl}

In this section we study the scattering of closed bosonic string states from a stack of D$p$-branes. In particular we consider disk and annulus amplitudes with external tachyonic and massless states. The vertex operators describing the emission of a closed string tachyon $V_T$ and the massless states\footnote{Although we shall consider the full set of massless states (graviton, dilaton and antisymmetric tensor) we shall often refer to the external state as the graviton.} $V_{\cal G}$ are
\begin{equation}
  \label{eq:clo-tac}
  V_T = \frac{\kappa_{26}}{2\pi} {\rm e}^{\ii p X}~,~~~ 
  V_{\cal G} = \frac{\kappa_{26}}{2\pi} \frac{2}{\alpha'} {\cal G}_{\mu \nu} \ii \partial X^\mu \, \ii \bar\partial X^\nu{\rm e}^{\ii p X},
\end{equation}
where $\kappa_{26}$ is the gravitational constant in the critical dimension for the bosonic theory ($d=26$).

The disk amplitude contributing to the scattering of a tachyon off a stack of $N$ coincident D$p$-branes is
\begin{equation}
  \label{eq:diam}
  {\cal A}^{TT}_1 = C_{S^2} \kappa_{26} \frac{\alpha'}{8\pi} N \int \frac{d^2z_1 d^2z_2}{dV_{abc}} \langle B |V_T(z_1) V_T(z_2) |0 \rangle~,
\end{equation}
where the boundary state $|B\rangle$ describes the D-branes and the sphere normalization $C_{S^2}$  is given by $C_{S^2}\left(\frac{\kappa_{26}}{2 \pi}\right)^2 \frac{\alpha'}{8\pi}=1$. Each insertion of a boundary state is accompanied by a factor of $\kappa_{26}$, a propagator~\eqref{eq:clP} and a factor of $N$. A brief summary of our conventions, of  the main steps in the evaluation 
of~\eqref{eq:diam} and of the other string amplitudes discussed in this section can be found 
in Appendix~\ref{AppA}, while here we list just the main results. From~\eqref{eq:diam} we obtain\footnote{We do not explicitly write the momentum conservation $(2\pi)^{p+1} \delta^{p+1}(p_1+p_2)$ common to all amplitudes.}
\begin{equation}
  \label{eq:ATT}
 {\cal A}^{TT}_1 = \frac{\kappa_{26} T_p}{2} N \frac{ \Gamma (-1-\alpha' s) \Gamma (-1 -\frac{\alpha'}{4} t)}{\Gamma (-2 -\alpha' s -  \frac{\alpha'}{4} t)}~,
\end{equation}
where $\sqrt{s}$ is the energy\footnote{We choose to work in a frame where the energy is the only non-zero component of the closed string momentum along the D-brane.} of the incident (or outgoing) closed string state $\sqrt{s}=p_1^0=|p_2^0|$ and $t$ is related to the momentum exchanged between the probe and the D$p$-branes: $t=-q^2 = -(p_1+p_2)^2$. In the Regge limit $s \gg |t|$ we have~\footnote{The combination $\frac{1}{2} \kappa_{26} T_p N$ is given in terms of the radius $R_p$ of the brane geometry in (\ref{Rp}).}
\begin{equation}
  \label{Atach}
  {\cal{A}}^{TT}_1 \sim \frac{\kappa_{26} T_p}{2} N
 {\rm e}^{-\ii \pi \frac{\alpha't}{4}} (\alpha' s)^{1 +\frac{\alpha't}{4}}  \frac{\Gamma \left(- \frac{\alpha' t }{4} \right)}{1+ \frac{\alpha't}{4} } \equiv A_1(s,q)~.
\end{equation}
Similarly the tree-level scattering of a massless state is given by the disk amplitude with two massless vertex operators
\begin{equation}
  \label{eq:diam2}
  {\cal A}^{{\cal G}{\cal G}}_1 = C_{S^2} \kappa_{26} \frac{\alpha'}{8\pi} N \int \frac{d^2z_1 d^2z_2}{dV_{abc}} \langle B |V_{\cal G}(z_1) V_{\cal G}(z_2) |0 \rangle~.
\end{equation}
The explicit expression of the full amplitude, when written in terms of the polarizations, is 
somewhat lengthy because several Lorentz structures are present. However it simplifies considerably in the Regge limit. As
discussed in detail in \cite{D'Appollonio:2013hja}, the simplest method to derive the asymptotic form of a generic string amplitude
is to use the Reggeon vertex \cite{Ademollo:1989ag, Ademollo:1990sd, Brower:2006ea}. For the amplitude \eqref{eq:diam2} the result is: 
\begin{eqnarray}
\label{calA1'}
{\cal{A}}_{1}^{{\cal G}{\cal G}} & \sim & - \frac{\kappa_{26} T_p}{2} N
 {\rm e}^{-\ii \pi \frac{\alpha't}{4}} (\alpha' s)^{1 +\frac{\alpha't}{4}}  \Gamma \left(- 1- \frac{\alpha' t }{4} \right) \\
 & \times & \left( (\epsilon_1 \epsilon_2)  - \frac{\alpha'}{2}  ({{\epsilon}}_1 q) ({{\epsilon}}_2 q)   \right)\left( ({\bar{\epsilon}}_1 {\bar{\epsilon}}_2) -
\frac{\alpha'}{2} ({\bar{\epsilon}}_1 q) ({\bar{\epsilon}}_2 q) \right) \nonumber ~,
\end{eqnarray}
where, as usual, we split the closed string polarization into its holomorphic and anti-holomorphic part ${\cal G}_{\mu\nu}=\epsilon_{\mu} \bar{\epsilon}_\nu$. Notice that the quantity in the first line in~(\ref{calA1'}) is just the elastic scattering of a tachyon on a D-brane~\eqref{Atach}.

This amplitude can be compared with the one for the type II superstring theories (see for instance Eq.~(2.12) of~\cite{D'Appollonio:2010ae})
\begin{eqnarray}
{\cal{A}}_{1}^{{\rm II}} \sim (\epsilon_1 \epsilon_2) ({\bar{\epsilon}}_1 {\bar{\epsilon}}_2) \frac{\kappa_{10} T_p^{\rm II}}{2} N \Gamma \left(- \frac{\alpha' t }{4} \right)  {\rm e}^{-\ii \pi \frac{\alpha't}{4}} (\alpha' s)^{1 +\frac{\alpha't}{4}} \equiv A_1^{\rm II}(s,q) ~.
\label{super'}
\end{eqnarray}
Apart from the overall factor which involves the 10D gravitational constant $\kappa_{10}$ and the boundary state normalization $T_p^{\rm II}$ (which is given by~\eqref{eq:Tp} for $d=10$), the two amplitudes have two main qualitative differences. 
The first is that the leading Regge trajectory includes the tachyonic (ground) state of
the bosonic theory.
The second is that even 
in the high energy Regge limit the bosonic amplitude has a non-trivial dependence on the polarization tensors, 
see the second line in~(\ref{calA1'}). 
As discussed in detail in~\cite{Camanho:2014apa}, this is a direct consequence of the modification of the three-graviton vertex in the bosonic theory which yields a quadratic ($\mbox{Riemann}^2$) and a cubic term ($\mbox{Riemann}^3$) in the effective action, while in the maximally supersymmetric case these corrections are forbidden by supersymmetry. Because of this, the Lorentz structure in~(\ref{calA1'}) is the same as the one appearing in~\cite{Camanho:2014apa}, see for instance Eq.~(3.17) therein\footnote{The only difference between the two setups is that in the tree-level graviton-brane scattering there is a single three-point vertex, while in the graviton-graviton scattering the same vertex appears twice (and so in~\cite{Camanho:2014apa}, the Lorentz structure in the second line in~(\ref{calA1'}) appears twice).}.

In the point-particle limit, the superstring result~\eqref{super'} becomes 
\begin{equation}
  \label{eq:sugraA1}
  {\cal A}_{1}^{{\rm II}} \to (2 E) 2 \delta(s,q) = (\epsilon_1 \epsilon_2) ({\bar{\epsilon}}_1 {\bar{\epsilon}}_2) \frac{\kappa_{10} T_p^{\rm II}}{2} N \frac{4}{q^2}s ~.
\end{equation}
We can construct the elastic $c$-number eikonal as the Fourier transform of $\delta(s,q)$ in the $d-(p+1)-1$ dimensional space (with $d=10$ for the superstring) transverse to the D$p$-brane and the direction of the incident state. In the superstring case the tidal excitations of the closed string are accounted for by replacing the $c$-number eikonal by an eikonal operator~\cite{Amati:1987uf} 
given by\footnote{From now on we will use the boldface to indicate the light-cone momenta. 
The physical polarizations of the massless states live naturally in $d-2$ space 
dimensions and the momentum exchanged $q$ has a vanishing small component along the direction of the incident state. See~\cite{D'Appollonio:2013hja} for a general discussion of the light-cone states also at the massive level.}
\begin{eqnarray}
2 {\hat{\delta}}^{\rm II} (s, {\mathbf b}) = \int_{0}^{2 \pi} \frac{d\sigma}{2\pi} \int \frac{d^{8-p} {\mathbf q}}{(2\pi)^{8-p}} \frac{A_1^{\rm II} (s, {\mathbf q})}{2E} : {\rm e}^{\ii \mathbf{q} (\mathbf{b}+ {\hat{X}} (\sigma))}: \ , 
\label{eikope}
\end{eqnarray}
where the operators
\begin{eqnarray}
{\hat{X}}^I (\sigma) = \ii \sqrt{\frac{\alpha'}{2}} \sum_{n=1}^{\infty} \left( \frac{a_n^I}{n} {\rm e}^{\ii n\sigma} - \frac{a_{-n}^I}{n} {\rm e}^{-\ii n\sigma} +  \frac{{\tilde{a}^I}_n}{n} {\rm e}^{-\ii n\sigma} - \frac{{\tilde{a}^I}_{-n}}{n} {\rm e}^{\ii n \sigma}  \right)\, \, , 
\hspace{0.2cm} I = 1, ..., d-2 \ , 
\label{hatX}
\end{eqnarray}
are the string coordinates in a light-cone gauge aligned to the collision axis, as discussed in~\cite{D'Appollonio:2013hja}.
A similar eikonal operator $2 \hat{\delta}(s,\mathbf{b})$ can be introduced in the bosonic case, simply by changing the critical dimension to $d=26$ and by using the amplitude $A_1$ in~\eqref{Atach}
\begin{eqnarray}
2 {\hat{\delta}} (s, {\mathbf b}) = \int_{0}^{2 \pi} \frac{d\sigma}{2\pi} \int \frac{d^{24-p} {\mathbf q}}{(2\pi)^{24-p}} \frac{A_1 (s, {\mathbf q})}{2E} : {\rm e}^{\ii \mathbf{q} (\mathbf{b}+ {\hat{X}} (\sigma))}:~.
\label{eikopeb}
\end{eqnarray}
By construction, the matrix element
of the operator $2 {\hat{\delta}} (s, \mathbf{b}) $ between two vacuum states of the Fock space of the $a^I_n$ in~\eqref{hatX}
gives the high-energy limit of the elastic scattering amplitude of a tachyon off a stack of D$p$-branes in the impact parameter representation
\begin{equation}
\langle 0| 2 {\hat{\delta}} (s, \mathbf{b}) | 0 \rangle =\int \frac{d^{24-p} \mathbf{q}}{(2\pi)^{24-p}} \frac{{A}_1 (s, \mathbf{q})}{2E} {\rm e}^{i \mathbf{q} \mathbf{b}}~.
\label{tach}
\end{equation}
Let us verify that the same holds in the case of two gravitons, which are naturally represented by  $|\epsilon,\bar\epsilon\rangle = \epsilon_I {\bar\epsilon}_{J} a_{-1}^I \tilde{a}_{-1}^J |0\rangle$
\begin{eqnarray}
\label{grav}
\langle \epsilon_1,\bar\epsilon_1 | 2 {\hat{\delta}} (s, \mathbf{b}) |\epsilon_2,\bar\epsilon_2\rangle & = & \langle \epsilon_1,\bar\epsilon_1 | \,\,2 {\hat{\delta}} (s, \mathbf{b}) \,\, |\epsilon_2,\bar\epsilon_2 \rangle 
\\ & = & \int \frac{d^{24-p} \mathbf{q}}{(2\pi)^{24-p}} \frac{{A}_1 (s, - \mathbf{q}^2)}{2E} {\rm e}^{i \mathbf{q}  \mathbf{b}} \, \int_{0}^{2 \pi} \frac{d\sigma}{2\pi} \, \langle \epsilon_1,\bar\epsilon_1 | : {\rm e}^{\ii \mathbf{q} {\hat{X}} (\sigma)}: |\epsilon_2,\bar\epsilon_2\rangle\,,
\nonumber 
\end{eqnarray}
where the last matrix element is equal to
\begin{align}
\label{eq:eeeem}
&\int_{0}^{2 \pi} \!\frac{d\sigma}{2\pi} \,  \langle \epsilon_1,\bar\epsilon_1 | \,  \left( 1 - \frac{1}{2} : ( \mathbf{q} {\hat{X}} (\sigma) )^2: + \frac{1}{24} : ( \mathbf{q} {\hat{X}} (\sigma) )^4: \right)\, |\epsilon_2,\bar\epsilon_2 \rangle = 
\\ \nonumber
& (\epsilon_1 \epsilon_2) ({\bar{\epsilon}}_1 {\bar{\epsilon}}_2)  - \frac{\alpha'}{2} (\mathbf{q} \epsilon_1)(\mathbf{q} \epsilon_2)  ({\bar{\epsilon}}_1 {\bar{\epsilon}}_2)  - \frac{\alpha'}{2} (\mathbf{q} {\bar{\epsilon}}_1)(\mathbf{q} {\bar{\epsilon}}_2) (\epsilon_1 \epsilon_2) \nonumber 
+ \frac{(\alpha')^2}{4}(\mathbf{q} {\bar{\epsilon}}_1)(\mathbf{q} {\bar{\epsilon}}_2) (\mathbf{q} \epsilon_1)(\mathbf{q} \epsilon_2) \,.
\end{align}
By using this result in~(\ref{grav}), one immediately obtains the Fourier transform of the amplitude in Eq.~(\ref{calA1'}) (apart from the usual factor of $2E$). 

Following~\cite{Black:2011ep,D'Appollonio:2013hja}, one can generalize this result and show that   
 the matrix elements of the 
eikonal operator $2 \hat\delta$~\eqref{eikopeb} between generic light-cone states of the bosonic string give
the Regge limit of a generic two-point function on the disk. This can be 
proved by studying the full set of inelastic scattering amplitudes at tree level, either using the
Reggeon vertex and the DDF operators or taking the Regge limit of the light-cone
three-string vertex.

It is also natural to assume that the bosonic eikonal operator $2{\hat{\delta}}$ exponentiates when one includes 
the leading contribution at high energy of the amplitudes with $h$ boundaries. As in the superstring case~\cite{Amati:1987uf},
one can argue that at the leading order in the Regge limit the string amplitudes with many boundaries are dominated by an integration 
region in the moduli space where the worldsheet looks like a (half) ladder diagram, with gravitons\footnote{Actually the whole leading Regge trajectory contributes at the string level.} connecting each boundary to the line representing the external energetic state. In formulae, this means that the leading contribution to tachyon scattering from the amplitude ${\cal A}^{TT}_h$ with $h$ boundaries should be a convolution of $h$ copies of the disk result~\eqref{Atach}
\begin{equation}
  \label{eq:conv}
  \frac{{\cal A}^{TT}_h}{2 E} \sim \frac{\ii^{h-1}}{h!} \int  \prod_{i=1}^{h} \frac{ d^{24-p} \mathbf{k_i}}{ (2\pi)^{24- p}}  \frac{A_1(s,\mathbf{k_i})}{2E} \langle 0| \prod_{i=1}^{h} \int\limits_0^{2 \pi} \frac{ d \sigma_i}{2 \pi} :{\rm e}^{\ii \mathbf{k}_i {\hat{X}}(\sigma_i)} : | 0 \rangle \delta\left(\sum \mathbf{k_i}- \mathbf{q}\right)\,.
\end{equation}
These convolutions become just products after the Fourier transform to the impact parameter space. 
Thus in the Regge limit the S-matrix takes the form 
\begin{eqnarray}
S = {\rm e}^{2 \ii \hat{\delta}(s,\mathbf{b})}~,
\label{exp}
\end{eqnarray}
which is obtained from~\eqref{eq:conv} by summing over $h$ and removing
 the two vacuum states that project the 
operator onto the ground (tachyonic) state. 
The eikonal operator in \eqref{exp} correctly describes the tidal effects 
(inelastic excitations) at the leading order in the small $R_p/b$ expansion.
This was discussed in the maximally supersymmetric context of the graviton-graviton scattering in~\cite{Amati:1987uf} and in~\cite{D'Appollonio:2010ae} for the string-brane case. A check of the validity of~(\ref{eq:conv}) in the bosonic case is given in Appendix~\ref{app:1loop} where the leading energy contribution to the two-tachyon amplitude at the one-loop (annulus) level is calculated.

\section{Causality problems in field theory}
\label{3}
\setcounter{equation}{0}

The results of the previous section show that   
the Regge limit of a generic two-point function on the disk is given by 
 the matrix element of the 
eikonal operator $2 \hat\delta$~\eqref{eikopeb} between the corresponding light-cone states.
Sewing together different copies of this operator one finds the leading contribution of the ladder diagrams. In this section we shall study the field theory truncation of the string amplitudes, restricting both the external and the intermediate states to the massless sector.

The projection of  $2 \hat\delta$ on the massless sector is given by~\eqref{grav}, with $\alpha'$ set to zero in the function $A_1$
\begin{equation}
  \label{eq:dihjk}
  2 \delta^{\rm ft}_{IH,JK}(E,\mathbf{q}) = \kappa_{26} T_p N \frac{E}{\mathbf{q}^2}
\left(\delta_{IJ} - \frac{\alpha'}{2} \mathbf{q}_I \mathbf{q}_J\right)
\left(\delta_{HK}  - \frac{\alpha'}{2} \mathbf{q}_H \mathbf{q}_K\right)~,
\end{equation}
and  describes the transition from the state $a_{-1}^I \tilde{a}_{-1}^H|0\rangle$ to 
the state $a_{-1}^J \tilde{a}_{-1}^K|0\rangle$. By gluing together $h$ copies of this operator with $h-1$ high-energy propagators one immediately obtains the field theory version of~\eqref{eq:conv}, where only massless fields propagate. This is most easily done after diagonalising the operator~\eqref{eq:dihjk}.
In the impact parameter space we have
\begin{equation}
2 \delta^{\rm ft}_{IH,JK} = K_0 
\delta_{IJ} \delta_{HK} + K_2 \left(\delta_{IJ} \Pi_{HK} +   \delta_{HK} \Pi_{IJ}\right)  + K_4 \Pi_{IJHK} \,,
\label{Amasslesssb}
\end{equation}
where 
\begin{equation}
K_0 = \frac{\kappa_{26} N T_p E \Gamma ( \frac{D-2}{2})}{ 4 \pi^{\frac{D}{2}} \mathbf{b}^{D-2} } ~,~~
K_2 = - \frac{\alpha'}{\mathbf{b}^2} \frac{\kappa_{26} N T_p E \Gamma ( \frac{D}{2})}{4 \pi^{\frac{D}{2}} \mathbf{b}^{D-2} }~,~~ K_4 = \left(\frac{\alpha'}{\mathbf{b}^2}\right)^2 \frac{\kappa_{26} N T_p E \Gamma ( \frac{D+2}{2})}{4\pi^{\frac{D}{2}} \mathbf{b}^{D-2} }~,
\label{masslessoperator}
\end{equation}
and the tensors $\Pi$ are trivial whenever one of the indices is along the $p$-dimensional space parallel to the D-branes, while in the $D = 24-p$ space where $\mathbf{b}$ lives are\footnote{When $p=22$, we have $D=2$ and the leading eikonal phase $K_0$ is proportional to $\ln \mathbf{b}^2$.}
\begin{align}
\label{PiPi}
\Pi_{ij} & = \left( \delta_{ij} - D \frac{\mathbf{b}_i \mathbf{b}_j }{\mathbf{b}^2} \right) \ ,
\\ \nonumber 
\Pi_{ijhk} & = \delta_{hk} \delta_{ij} +
\delta_{hj} \delta_{ik} + \delta_{jk} \delta_{ih} - \frac{D+2}{\mathbf{b}^2}\left(\mathbf{b}_h \mathbf{b}_k \delta_{ij} + \mathbf{b}_h \mathbf{b}_j \delta_{ik} {{{} \atop {}} \atop {{} \atop {}}}  \right. \\  \nonumber 
& \left.  + \, \mathbf{b}_i \mathbf{b}_h \delta_{jk} + \mathbf{b}_j \mathbf{b}_k \delta_{ih} + \mathbf{b}_i \mathbf{b}_k \delta_{jh} + \mathbf{b}_i \mathbf{b}_j \delta_{hk} -
(D+4) \frac{\mathbf{b}_i \mathbf{b}_j \mathbf{b}_h \mathbf{b}_k}{\mathbf{b}^2}\right) \ .
\end{align}
By constructions the $\Pi$'s are completely symmetric and traceless in all indices since they are obtained by taking derivative of $1/\mathbf{b}^{D-2}$,  the Green function of the Laplacian in the impact parameter space. 

It is then straightforward to check that the massless states split into several groups with distinct eigenvalues (we are subtracting 
below the common term $K_0$ present in all the eigenvalues)
\begin{itemize}
\item $p^2$ states of eigenvalue zero corresponding to the metric and $B$-field polarizations ${\cal G}_{a_1 a_2}$, where $a_1,a_2$ are along the $p$ space directions of the D$p$-branes.
\item  $2 p (D-1)$ states of eigenvalue $K_2$ corresponding to the metric and $B$-field polarizations ${\cal G}_{a_1 \,\hat{\! j}}$, where the hatted indices, such as $\hat{\! j}$, are perpedicular to both the D-branes and the impact parameter vector $\mathbf{b}$.
\item $2 p$ states of eigenvalue $(1-D) K_2$ corresponding to the metric and $B$-field polarizations ${\cal G}_{a_1 \widehat{\mathbf{b}}}$, where $\widehat{\mathbf{b}}$ indicates the direction along the impact parameter.
\item $(D-1)(D-2)/2$ states of eigenvalue $2 K_2$ corresponding to the polarizations $B_{\,\hat{\! i}\,\hat{\! j}}$ of the $B$ field.
\item $(D-1)$ states of eigenvalue $(2-D) K_2$ corresponding to the polarizations $B_{\,\hat{\! i} \,\widehat{\mathbf b}}$.
\item $(D-1)(D-2)/2$ states of eigenvalue $2 K_2 + 2 K_4$  corresponding to the metric polarizations $G_{\,\hat{\! i} \,\hat{\! j}}$ with ${\,\hat{\! i}} \not = {\,\hat{\! j}}$.
\item $(D-1)$ states of eigenvalue $(2-D) K_2 -2(D+1) K_4$ corresponding to the metric polarizations $G_{\,\hat{\! i} \widehat{\mathbf{b}}}$.
\item $(D-2)$ states of eigenvalue $2 K_2 +2 K_4$ corresponding to the  metric polarizations  $G_{\,\hat{\! i} \,\hat{\! i}} - G_{\,\hat{\! i}+1 \,\hat{\! i}+1}$, where $\,\hat{\! i}$ can take only the first $D-2$ values.
\item The remaining two eigenstates are a mixture of the states $|v_1\rangle = {\cal G}_{\,\hat{\! i} \,\hat{\! j}} \delta^{\,\hat{\! i} \,\hat{\! j}}/\sqrt{D-1}$ and $|v_2 \rangle = {\cal G}_{\widehat{\mathbf{b}} \widehat{\mathbf{b}}}$. In the subspace generated by these two vectors the shift in the eikonal operator can be represented by a $2\times 2$ matrix $M^{(2)}$ whose elements are $M^{(2)}_{11}=a$, $M^{(2)}_{22} =b$, $M^{(2)}_{12}=M^{(2)}_{21}=c$, where
  \begin{equation}
    \label{eq:abc}
a = 2 K_2 + (D+1) K_4\;, ~
b = -2 (D-1) K_2 +  (D^2-1) K_4\;,~
c = - (D+1) \sqrt{D-1} K_4\;.
  \end{equation}
The eigenvalues $\lambda_{\pm}$ and the eigenstates $|v_{\pm}\rangle$ of the matrix $M^{(2)}$ 
turn out to be
\begin{equation}
\label{eigen}
\lambda_{\pm} = \frac{a+b}{2} \pm \frac{1}{2} \sqrt{(a-b)^2+ 4 c^2}\;,~~
|v_{\pm}\rangle = \frac{(\lambda_{\pm}-b) |v_1\rangle + {c} |v_2\rangle}{\sqrt{ (\lambda_{\pm}-b)^2 + c^2}}\;.
\end{equation}
\end{itemize}
The states listed above diagonalize the $S$-matrix in the Born approximation. It seems reasonable to assume that the most general interaction describing the scattering of massless states off a heavy target and consistent with the assumptions mentioned after~\eqref{higherder} has the same form of~\eqref{eq:dihjk}, but with arbitrary coefficients for the ${\mathbf q}^2$ and ${\mathbf q}^4$ terms related to $l_2$ and $l_4$ in~\eqref{higherder}. In this basis it is easy to resum the half-ladder diagrams constructed by gluing together many vertices of this type together with the high energy propagator $1/(2E)$: one obtains a convolution in momentum space and so a power of vertex~\eqref{Amasslesssb} in the impact parameter space. In the eigenstate basis then the $S$-matrix is elastic and the half-ladder diagrams reconstruct the exponential $S = \exp(\ii \lambda_O)$ for each eigenvalue $\lambda_O$.

For each eigenstate, deflection angle\footnote{Following~\cite{Camanho:2014apa}, in order
to avoid the deflection of the probe particle one could discuss  the propagation
of  a closed string between two equidistant stacks of D$p$-branes.}
 and time delay \cite{Ciafaloni:2014esa}, \cite{Camanho:2014apa} can be computed by the appropriate derivatives of the  corresponding
eikonal phase. It is clear that the above results lead to causality violations in some particular channels, independently of the particular expressions we have for $K_2$ and $K_4$.
Causality violations arise for instance for the components $G_{a_1 \hat j}$
and $G_{\hat i \hat {\bf b}}$ of the metric and for the components
$B_{a_1 \hat j}$ and $B_{\hat i \hat j}$ of the Kalb-Ramond field.
We conclude that our truncation of string theory leads to arbitrarily large causality violations for sufficiently small values of $\frac{l_{2,4}}{b}$ (but still with $b > R_p$ and $R_p$ sufficiently small).

From the result above it is clear that the dilaton, defined as the trace in the light-cone space  $|w_1\rangle = ({\cal G}_{a_1 a_2} \delta^{a_1 a_2} + {\cal G}_{\,\hat{\! i} \,\hat{\! j}} \delta^{\,\hat{\! i} \,\hat{\! j}} + {\cal G}_{\widehat{\mathbf{b}} \widehat{\mathbf{b}}})/\sqrt{D+p}$, is coupled to the metric components involving the diagonal elements ${\cal G}_{\,\hat{\! i} \,\hat{\! i}}$ and ${\cal G}_{a_1 a_1}$. We can make this explicit by starting from the eikonal matrix in the space generated by $|v_0\rangle = {\cal G}_{a_1 a_2} \delta^{a_1 a_2}/\sqrt{p}$, $|v_1\rangle$ and $|v_2\rangle$, and rewriting it in a basis defined by the dilaton $|w_1\rangle$ and other two vectors, $|w_0\rangle$ and $|w_2\rangle$, representing pure metric fluctuations
\begin{align}
\nonumber
  |w_0\rangle & = \sqrt{\frac{D}{D+p}} |v_0\rangle - \sqrt{\frac{(D-1) p}{D(D+p)}} |v_1\rangle - \sqrt{\frac{p}{D (D+p)}} |v_2\rangle ~, \\
  |w_1\rangle & = \sqrt{\frac{p}{D+p}} |v_0\rangle + \sqrt{\frac{D-1}{D+p}} |v_1\rangle + \frac{1}{\sqrt{D+p}} |v_2\rangle~,   \label{eq:w} \\
  |w_2\rangle & = - \frac{1}{\sqrt{D}} |v_1 \rangle + \sqrt{\frac{D-1}{D}} |v_2\rangle~. \nonumber
\end{align}
We can decouple the dilaton, so as to obtain the result for pure gravity, as follows: we consider the $3 \times 3$ matrix $M^{(3)}$ obtained by adding a row and a column of zeros representing the state $|v_0\rangle$, then we write the eikonal phase in the new $|w_i\rangle$ basis~\eqref{eq:w}, and finally we eliminate the second row and column that correspond to the dilaton $|w_1\rangle$. The eigenvalues of the new $2\times 2$ matrix obtained in this way correpond to the pure gravity eigenstates in the space spanned by\footnote{
 This procedure allows one to check that for $p=0, D =2$ (an unphysical case for the bosonic string since $D+p$ should be $24$) there is no correction proportional to $K_2$. This follows from the fact that, in the case of pure gravity, the $R^2$ correction is a total derivative (Gauss-Bonnet) in four-dimensions \cite{Camanho:2014apa}.} $|w_0\rangle$ and $|w_2\rangle$.

We conclude this Section with a comment on the maximally supersymmetric case. 
Although in Type II supergravity the graviton three-point coupling does not receive any correction,
the causality problem identified in~\cite{Camanho:2014apa} can arise 
 if one considers the propagation of a higher spin massive particle. 
For instance the first massive level of the superstring spectrum
contains a class of states that transform in the
tensor product of two totally symmetric traceless tensors of rank two. Their scattering amplitudes on a
stack of D-branes in the Regge limit can be evaluated using the methods
introduced in~\cite{D'Appollonio:2013hja}. For states polarized in the
directions transverse to the branes and to the collision axis the result is
proportional to a polynomial in the polarizations and the momentum transfer
which is essentially the same as the one in \eqref{eq:eeeem}
\ba &&
(\epsilon_{1, ij} \epsilon_2^{ij}) ({\bar{\epsilon}}_{1, kl} {\bar{\epsilon}}_2^{kl})  
- \frac{\alpha'}{2} (\epsilon_{1, ij}\mathbf{q}^j )(\epsilon_{2}{}^i{}_s \mathbf{q}^s )  
({\bar{\epsilon}}_{1, kl} {\bar{\epsilon}}_2^{kl})   
- \frac{\alpha'}{2}(\epsilon_{1, ij} \epsilon_2^{ij})
( {\bar{\epsilon}}_{1, kl}\mathbf{q}^l )( {\bar{\epsilon}}_2{}^k{}_r\mathbf{q}^r )   
\nonumber \\ 
&+& \frac{(\alpha')^2}{4} (\epsilon_{1, ij}\mathbf{q}^j )(\epsilon_{2}{}^i{}_s \mathbf{q}^s )  
( {\bar{\epsilon}}_{1, kl}\mathbf{q}^l )( {\bar{\epsilon}}_2{}^k{}_r\mathbf{q}^r )    \ ,
\ea
where
we split the rank-four polarization tensor
into its holomorphic and anti-holomorphic parts ${\cal S}_{\mu\nu\rho\sigma}=\epsilon_{\mu\nu} \bar{\epsilon}_{\rho\sigma}$. 
Therefore if we were to allow only the propagation of massless states in the transverse channel 
or even if we were to include a finite number of higher spin massive fields, a causality problem
would arise. This is not as compelling as the one in~\cite{Camanho:2014apa} because
 a simple field theory description of the consistent interactions 
of a massive higher spin particle with the gravitational field is still lacking and therefore
we are forced to work in the context of string theory  from the very beginning.
It is however interesting to note that generic non-minimal  couplings
of a higher spin field to the graviton would lead to negative Shapiro time delays.

\section{String theory resolution of the causality problems}
\label{4}
\setcounter{equation}{0}

In this Section we study how the causality problem that arises in the field 
theory limit is solved when one takes into account the complete string dynamics. 
Paralleling the treatment in \cite{Amati:1987uf} let us start with the superstring case where, in the Regge limit,
\begin{eqnarray}
\frac{{\cal{A}}_1 (s, - {\mathbf q}^2)}{2E} =  \frac{1}{2} N T_p \kappa_{10} \, \Gamma \left( - \frac{\alpha' t}{4}\right)  \frac{(\alpha' s)}{2E}{\rm e}^{  \frac{\alpha' t}{4} \bar{Y}}\ , \hspace{0.4cm} \bar{Y} = Y - \ii\pi \equiv \log (\alpha' s) - \ii \pi \ . 
\label{calAsuper}
\end{eqnarray}
In impact parameter space we find
\begin{eqnarray}
&& \frac{{\cal{A}}_1 ( s,{\mathbf b}) }{2E} \equiv \int \frac{d^{8-p} {\mathbf q}}{(2\pi)^{8-p}} \frac{{\cal{A}}_{1} (s, t)}{2E} {\rm e}^{\ii {\mathbf q} {\mathbf b}} = 
\frac{1}{2} N T_p \kappa_{10} \, \Gamma \left ( 1 + \frac{\alpha'}{4} \nabla^2 \right )  \nonumber \\
&& \times  \int  \frac{d^{8-p} {\mathbf q}}{(2\pi)^{8-p}}  \frac{ 4 s}{2E {\mathbf q}^2} \exp \left(- \frac{\alpha'}{4} {\mathbf q}^2 \bar{Y} + \ii {\mathbf q} {\mathbf b}   \right) \ , 
\hspace{0.4cm} t = - {\mathbf q}^2 = \frac{\partial^2}{\partial {\mathbf b}^i \partial {\mathbf b}^i} \equiv \nabla^2 \; \ .
\label{calA1bc}
\end{eqnarray}
We can  compute the integral in the second line as follows
\begin{eqnarray}
&& \frac{4s}{2E} \int_0^{\infty} dT \int  \frac{d^{8-p} {\mathbf q}}{(2\pi)^{8-p}}  {\rm e}^{- {\mathbf q}^2 \left(T + \frac{\alpha'}{4}  \bar{Y}    \right) + \ii {\mathbf q} {\mathbf b} } = \frac{4s}{(4\pi)^{\frac{8-p}{2}}}  \int_0^{\infty} dT  \left(T + \frac{\alpha'}{4} \bar{Y} \right)^{\frac{p-8}{2}}  {\rm e}^{ - \frac{{\mathbf b}^2}{4\left( T + \frac{\alpha'}{4}  \bar{Y} \right)}} \nonumber \\
&&  =  \frac{2E}{(4\pi)^{\frac{8-p}{2}}}    \int_{\frac{\alpha'}{4}  \bar{Y} }^{\infty} d \hat{T} \hat{T}^{\frac{p-8}{2}} {\rm e}^{- \frac{{\mathbf b}^2}{4 \hat{T} }} =   \frac{2E}{(4\pi)^{\frac{8-p}{2}}}  \left( \frac{{\mathbf b}^2}{4} \right)^{\frac{p}{2} -3}\int_{0}^{ \frac{{\mathbf b}^2}{\alpha' \bar{Y}}} d t \,\, t^{2 - \frac{p}{2}} {\rm e}^{-t}\; \ . 
\label{impactb}
\end{eqnarray}
The last integral gives an incomplete gamma-function
\begin{eqnarray}
\gamma (s;x) = \int_0^x dt t^{s-1} {\rm e}^{-t} = \sum_{k=0}^{\infty} \frac{x^{s+k} {\rm e}^{-x}}{s(s+1) \dots (s+k)} \Longrightarrow \frac{x^s}{s} + \dots \hspace{1cm}  {\rm for} \ x \ll 1 \ ,
\label{gamainco}
\end{eqnarray}
and therefore
\begin{eqnarray}
\frac{{\cal{A}}_1 ( s,{\mathbf b}) }{2E} = \frac{1}{2} N T_p \kappa_{10} \, \Gamma \left( 1 + \frac{\alpha'}{4} \nabla^2\right)   
\frac{2E}{(4\pi)^{\frac{8-p}{2}}}   \left( \frac{{\mathbf b}^2}{4} \right)^{\frac{p}{2} -3} \gamma \left( 3 - \frac{p}{2} ;  \frac{{\mathbf b}^2}{\alpha'\bar{Y}} \right) \ .
\label{ReA1}
\end{eqnarray}
At high energy and small impact parameter $b \ll \sqrt{\alpha'} Y$
\begin{eqnarray}
\frac{{\cal{A}}_1 (s,{\mathbf b}) }{2E}  \sim  \frac{1}{2} N T_p \kappa_{10} \, \Gamma \left( 1 + \frac{\alpha'}{4} \nabla^2\right)
  \frac{2E}{(4\pi)^{\frac{8-p}{2}}}  
\times \left( \frac{4}{\alpha'\bar{Y}} \right)^{3 - \frac{p}{2}} \frac{1 }{3 - \frac{p}{2}}
\left( 1+ O \left (  \frac{{\mathbf b}^2}{\alpha' \bar{Y}} \right)\right) \ . 
\label{lowb}
\end{eqnarray}
For the bosonic string we start from (\ref{calA1'}). Neglecting for the moment the polarization dependent prefactor
and proceeding exactly as before, we obtain instead of (\ref{ReA1})
\begin{eqnarray}
 \frac{{\cal{A}}_1 ( s,{\mathbf b}) }{2E}  =   \frac{1}{2} N T_p \kappa_{26} \, 
\frac{ \Gamma \left( 1 + \frac{\alpha'}{4} \nabla^2\right)}{1 - \frac{\alpha' }{4} \nabla^2}    \frac{2E}{(4\pi)^{\frac{24-p}{2}}}  
\left( \frac{{\mathbf b}^2}{4} \right)^{\frac{p}{2} -11} \gamma \left( 11 - \frac{p}{2} ;  \frac{{\mathbf b}^2}{\alpha' \bar{Y}} \right)\; \ , 
\label{cal1bos}
\end{eqnarray}
which, for ${\mathbf b} \ll \sqrt{\alpha'} Y$ and at high energy, becomes
\begin{eqnarray}
\frac{{\cal{A}}_1 ( s,{\mathbf b}) }{2E}  \sim  \frac{1}{2} N T_p \kappa_{26}\, \frac{ \Gamma \left( 1 + \frac{\alpha'}{4} \nabla^2\right)}{1 - \frac{\alpha' }{4} \nabla^2}  \frac{2E}{(4\pi)^{\frac{24-p}{2}}} 
 \left( \frac{4}{\alpha'\bar{Y}} \right)^{11 - \frac{p}{2}} \frac{1} {11 - \frac{p}{2}}
\left( 1+ O \left (  \frac{{\mathbf b}^2}{\alpha' \bar{Y}} \right)\right)  \ . 
\label{smallbbos}
\end{eqnarray}
The above tree-level scattering amplitudes in impact parameter space are expected to exponentiate in an operator form when higher loop corrections are included. This is checked to be the case at the annulus level in Appendix B.
The elastic scattering amplitude will be suppressed both by the imaginary part contained in (\ref{lowb})
and (\ref{smallbbos}) (through $\bar{Y}$) and by the replacement of the c-number eikonal by an operator eikonal. The former effect is connected to the possibility of producing open strings living on the brane system and will be discussed in a forthcoming paper \cite{DDRVopenclosed}. The latter phenomenon is related to tidal-force excitation of the incoming closed string, a process already discussed in detail in \cite{D'Appollonio:2013hja}.

For the purpose of this paper we ignore these absorptive effects and concentrate our attention on the real part of the c-number eikonal where we replace $\bar{Y}$ by $Y \equiv {\rm log}(\alpha' s)$. If we further notice that the operator $\alpha' \nabla^2$  acts on a function of $ x \equiv \frac{{\mathbf b}^2}{\alpha' Y}$, we see that, effectively, $\alpha' \nabla^2 \sim Y^{-1}\partial_x^2$. Such an observation allows us to approximate
at large $Y$ the differential operators appearing in  (\ref{lowb}) and (\ref{smallbbos}) with the identity operator.
In particular, for the bosonic string even the tachyonic pole at $ \nabla^2 = \frac{4}{\alpha' }$ becomes harmless\footnote{This is essentially due to the fact that tachyon exchange is suppressed 
by two powers of the energy with respect to graviton exchange and therefore it is negligible in the high-energy limit.} and we simply obtain
\begin{eqnarray}
\frac{Re\left({\cal{A}}_1 (s,{\mathbf b}) \right)}{2E}  \sim   \frac{1}{2} N T_p \kappa_{26} \,  \frac{2E}{(4\pi)^{\frac{24-p}{2}}} 
 \left( \frac{4}{\alpha' Y} \right)^{11 - \frac{p}{2}} \frac{1} {11 - \frac{p}{2}}
\left( 1+ O  \left(  \frac{{\mathbf b}^2}{\alpha'  Y} \right) \right)  \ . 
\label{smallbboslargeY}
\end{eqnarray}
The eikonal phase for the elastic scattering of a graviton can be obtained by acting on
the previous expression with the Fourier transform of a polynomial in the momenta and the polarizations.
As shown in Section \ref{strampl}, for the bosonic string this polynomial is
\be \left( (\epsilon_1 \epsilon_2)  - \frac{\alpha'}{2}  ({{\epsilon}}_1 {\mathbf q}) ({{\epsilon}}_2 {\mathbf q})   \right)\left( ({\bar{\epsilon}}_1 {\bar{\epsilon}}_2) - \frac{\alpha'}{2} ({\bar{\epsilon}}_1 {\mathbf q}) ({\bar{\epsilon}}_2 {\mathbf q}) \right) \ . \ee
In sharp contrast with the QFT limit discussed
in the previous Section, terms containing the momentum transfer ${\mathbf q}$ (which in the impact parameter space corresponds to
$\frac{\partial}{\partial {\mathbf b}}$) are parametrically small at high energy, being
suppressed by inverse powers of $Y$ with respect to terms of the same order in $|{\mathbf b}|$. Since the leading (Einstein-Hilbert) term respects causality, the  Shapiro time delay is positive for all
possible choices of the polarizations of the graviton, the Kalb-Ramond form and the dilaton.

At this point we could ask how essential   is the Regge behavior of the string amplitudes 
for the resolution of the causality problem. We could for instance include all the string corrections except for those related to 
Reggeization. Since 
the appearance of $Y$ in the transverse momentum cut off (and therefore in its ${\mathbf b}$-transform) is due to Regge behavior, we
could consider what happens if we  replace $Y$ by a constant. It is easy to see that this would not be sufficient to eliminate completely the danger of causality violations since the corrections to the EH time delay are of order one.
This is much better than the QFT situation discussed in the previous Section but not as good as 
the result obtained in this Section by taking into account the full dynamics of string theory. 


\section{Conclusions}
\label{Conclusions}
\setcounter{equation}{0}

The main aim of  this paper was to  illustrate, in the context of string theory, a recent observation by Camanho et al.~\cite{Camanho:2014apa} that adding to the Einstein-Hilbert action higher derivative corrections at scales lower than the Planck energy leads generically to short-distance violations of causality via negative time delays. To this purpose, we considered the high-energy collision of the massless modes  (graviton, dilaton and antisymmetric tensor)
of perturbative closed bosonic string theory (where such corrections to the EH action are present)  off a stack of 
a large number of coincident D$p$-branes. 

We restricted ourselves to the small-deflection-angle regime (corresponding to impact parameters $b$ much larger than the gravitational radius $R_p$ induced by the branes) for which string corrections are under control (see e.g.~\cite{D'Appollonio:2013hja} and references therein) even if $b < l_s = \sqrt{\alpha'}$.
Within this regime we considered different truncations of the full theory and checked whether the potential causality problems pointed out 
in~\cite{Camanho:2014apa}  appear  at each truncation level.

At the most drastic level, corresponding to a QFT limit with higher derivative terms in the effective action, we find that causality violations do emerge in certain channels defined by the relative orientation of the projectile polarization and the impact parameter vector. In this case the violations, corresponding to an increasingly negative time delay, become parametrically large (with respect to the positive Shapiro time delay) as $b$ becomes increasingly smaller than $\sqrt{\alpha'}$.

Within a somewhat milder truncation, which just replaces Regge behavior $s^{1 + \alpha' t/4}$ by a QFT-type fixed-power $s$, we find that causality-violating terms are of the same order as the causality-preserving ones. Generically, causality violations of order one 
will then occur in some channels. Finally, when the full Regge behavior is taken into account, the causality violation terms are suppressed by inverse powers of $Y = \log(\alpha's)$ and no violation occurs, as already pointed out in~\cite{Camanho:2014apa}.

The main lesson to be drawn from our exercise is the apparent necessity of Regge behavior for the resolution of the causality problem raised by CEMZ. Since Regge behavior needs the contribution of an infinite number of $t$-channel partial waves, this conclusion goes along very much with the one made in~\cite{Camanho:2014apa}  where arguments were given for the necessity of an infinite number of states of arbitrarily high spin exchanged in the $t$-channel. Although in a generic Quantum Field Theory Regge behavior can be obtained without invoking such a spectrum of single-particle states, their necessity looks inescapable in the weak coupling situation we have considered where $l_{2,4} \gg R_p \gg l_P$. 

Note finally that Regge behavior automatically induces a new phenomenon related to the branch cut  associated with a non-integer power of $s$  (which disappears in the QFT limit). Its correct interpretation  in string theory is well known from the very early days of the dual resonance model: it corresponds to the production of very massive strings in the $s$-channel. For string-string collisions these are heavy closed strings (discussed, for instance, in~\cite{Amati:1987uf} and~\cite{Veneziano:2004er}) while for the process at hand they are  open strings attached to the branes. In other words, at impact parameters smaller than the string length, the elastic amplitude is strongly reduced in favor of converting the (very high) energy of the massless projectile into (very) massive open strings.
A detailed analysis of this process will be discussed in a forthcoming paper~\cite{DDRVopenclosed}. 

\vspace{7mm}
\noindent {\large \textbf{Acknowledgements} }\\

GV would like to acknowledge discussions with Juan Maldacena and Adam Schwimmer and to thank the CCPP of NYU and the Physics department of Tel Aviv University for hospitality while part of this work was carried out. PDV thanks M. Bill{\'{o}}, M. Frau, A. Lerda and I. Pesando for many useful discussions on disk amplitudes. This work was partially supported by the Science and Technology Facilities Council Consolidated Grant ST/L000415/1 {\em ``String theory, gauge theory \&
duality''}.

\begin{appendix}

\section{Tree-level high-energy string-brane scattering}
\label{AppA}
\setcounter{equation}{0}

In this Appendix we provide some details about the disk amplitudes discussed in Section~\ref{strampl}. Beside the vertex operators the other main ingredient appearing in the disk amplitudes is the boundary state $|B\rangle$ 
\begin{equation}
  \label{eq:bs}
  |B\rangle = \frac{T_p}{2}\, \delta^{(d-1-p)} ( {\hat{q}}) \,\,\prod_{n=1}^{\infty}  {\rm e}^{- \alpha_{-n} R \,\,{\bar{\alpha}}_{-n}  } ~|0, p=0\rangle~,
\end{equation}
where $R$ is the diagonal reflection matrix,
whose elements are 
$R^\mu_{\;\nu} = \delta^{\mu}_{\;\nu}$ when $\mu,\nu=0,\ldots,p$ and $R^\mu_{\;\nu} = -\delta^{\mu}_{\;\nu}$ when $\mu,\nu=p+1,\ldots,26$. The string modes are defined as usual
\begin{equation}
  \label{eq:Xmod}
  X^\mu(z,\bar{z}) = q^\mu - \ii \frac{\alpha'}{2} p^\mu \ln|z|^2 + \ii \sqrt{\frac{\alpha'}{2}} \sum_{n\not= 0}^{\infty} \left( \frac{\alpha_n^\m}{n} z^{-n} -  \frac{{\tilde{\alpha}^\m}_n}{n} \bar{z}^{-n} \right)~.
\end{equation}
Finally, the D$p$-brane tension $\tau_p = T_p/\kappa_{26}$ in the bosonic theory is obtained from 
$\kappa_{26}$, given in (\ref{kappa2}), and from  
\begin{equation}
  \label{eq:Tp}
  T_p = \frac{\sqrt{\pi}}{2^{\frac{d-10}{4}}} (4 \pi^2 \alpha')^{ \frac{d-2p -4}{4}}~,~~\mbox{with}~~~ d=26~.
\end{equation}
Starting from~\eqref{eq:diam}, we can use the boundary state to transform all the right moving oscillators $\tilde{\alpha}_{-n}$ into left moving oscillators and obtain a correlator of four open-string-like vertices located in $1/\bar{z}_2$, $1/\bar{z}_1$, $z_1$, and $z_2$. 
Due to the residual $SL(2,R)$ invariance, we can fix three of these variables and then use the unintegrated form for
the corresponding vertices.
As usual $dV_{abc}$ is proportional to the ghost correlator from the unintegrated vertices. Introducing the cross-ratio
\begin{equation}
  \label{eq:crrat}
  x = \frac{\left(\frac{1}{\bar{z}_1} - z_1 \right) \left(\frac{1}{\bar{z}_2} - z_2 \right)} { \left(\frac{1}{\bar{z}_2} - z_1\right) \left(\frac{1}{\bar{z}_1} - z_2\right)}~,
\end{equation}
we have
\begin{equation}
  \label{eq:meas0}
  \frac{d^2z_1 d^2z_2}{dV_{abc}} = \left(\frac{1}{\bar{z}_1} - z_2 \right)^2 \left(\frac{1}{\bar{z}_2} -  z_1\right)^2\, dx~.
\end{equation}
Then the disk amplitude with two closed string tachyons~\eqref{eq:diam} is
\begin{eqnarray}
\label{eq:TTd1}
{\cal A}^{TT}_1 &=&  \frac{\kappa_{26} T_p}{2} N \int_0^1 dx 
\left(\frac{1}{\bar{z}_1} - z_2 \right)^2 \left(\frac{1}{\bar{z}_2} -  z_1\right)^2\left[
\left( \frac{1}{\bar{z}_1} - z_1 \right)^{ \frac{\alpha'}{2} p_1 R p_1  } 
\left( \frac{1}{\bar{z}_1} - z_2 \right)^{\frac{\alpha'}{2} p_1 R p_2  }\right. \nonumber \\
&  & \left.
\left(\frac{1}{\bar{z}_2} - \frac{1}{\bar{z}_1}\right)^{\frac{\alpha'}{2} p_1  p_2  }  
\left(z_1 - z_2\right)^{\frac{\alpha'}{2} p_1  p_2  } 
\left(\frac{1}{\bar{z}_2} - z_1 \right)^{\frac{\alpha'}{2} p_1 R p_2  }
\left(\frac{1}{\bar{z}_2} - z_2 \right)^{\frac{\alpha'}{2} p_2 R p_2  } \right] \\ \nonumber
 & =&  \frac{\kappa_{26} T_p}{2} N \int_0^1 dx \,\,x^{-\alpha' s-2} (1-x)^{-\frac{\alpha'}{4} t -2} ~,
\end{eqnarray}
where we used $C_{S^2}\left(\frac{\kappa_{26}}{2 \pi}\right)^2 \frac{\alpha'}{8\pi}=1$ and 
\begin{equation}
  \label{eq:pqst}
  2 p_1 p_2 = - t + m_1^2 + m_2^2 \ , \hspace{1cm} 
p_r R p_r = - 2s + m_r^2 \ , \hspace{1cm} 
2 p_1 R p_2 = 4s +  t - m_1^2 - m_2^2  ~,
\end{equation}
with $m_1^2=m_2^2=-4/\alpha'$. Then from~\eqref{eq:TTd1} we immediately obtain~\eqref{eq:ATT}.

The amplitude for the elastic scattering of a graviton on a disk is given by \eqref{eq:diam2}. The full amplitude can be written in terms of three 
types of integrals over the cross ratio $x$, depending on how many $\partial X$ operators are contracted

\begin{equation}
  \label{eq:diam2x}
  {\cal A}^{{\cal G}{\cal G}}_1 = \frac{\kappa_{26} T_p}{2} N \left(I_1 + I_2 + I_3\right)~.
\end{equation}
By using~\eqref{eq:pqst} with $m_1^2=m_2^2=0$, after a straightforward though somewhat long calculation one finds
the explicit form of the integrand in~\eqref{eq:diam2x}. The first contribution comes from the contraction of all 
the $\partial X$ operators among themselves
\begin{equation}
I_1 = \int_0^1 dx \,\,x^{-\alpha' s} (1-x)^{-\frac{\alpha'}{4} t}
 \left[ \frac{(\epsilon_1 \epsilon_2)   ({\bar{\epsilon}}_1 {\bar{\epsilon_2}})}{(1-x)^2}  + \frac{1}{x^2} 
  (\epsilon_1R {\bar{\epsilon}}_1)   ({\bar{\epsilon}}_1  R {{\epsilon_2}}) +
   (\epsilon_1R {\bar{\epsilon}}_2)   ({\bar{\epsilon}}_2  R {{\epsilon_1}}) 
   \right] ~.
\end{equation}
The term quadratic in the external momenta is
\begin{eqnarray}
\label{2mombis}
I_2 &=& \frac{\alpha'}{2} \int_0^1 dx \,\,x^{-\alpha' s} (1-x)^{-\frac{\alpha'}{4} t } 
 \\
&& \left[ - \frac{1-x}{x^2}  (\epsilon_1 R {\bar{\epsilon}}_1)  \epsilon_{2\mu} 
  \left( 2E \delta^{\mu}_{\,\,0} + q^{\mu} \frac{x}{1-x} \right){\bar{\epsilon}}_{2\nu} 
\left( 2E \delta^{\nu}_{\,\,0} + q^{\nu} \frac{x}{1-x} \right) \right. \nonumber \\
 && + \frac{1}{x} (\epsilon_1  \epsilon_2)  
  {\bar{\epsilon}}_{1\mu}  \left( 2E \delta^{\mu}_{\,\,0} - q^{\mu} \frac{x}{1-x} \right) 
   {\bar{\epsilon}}_{2\nu}  \left( 2E \delta^{\nu}_{\,\,0} + q^{\nu} \frac{x}{1-x} \right) \nonumber \\
 && - \frac{1-x}{x}  (\epsilon_1 R {\bar{\epsilon}}_2)   {\bar{\epsilon}}_{1\mu}
  \left( 2E \delta^{\mu}_{\,\,0} - q^{\mu} \frac{x}{1-x} \right)
 \epsilon_{2\nu}   \left( 2E \delta^{\nu}_{\,\,0} + q^{\nu} \frac{x}{1-x} \right)
 \nonumber \\
&&- \frac{1-x}{x}  ({\bar{\epsilon}}_1 R {{\epsilon}}_2)  \epsilon_{1\mu}
\left( 2E \delta^{\mu}_{\,\,0} - q^{\mu} \frac{x}{1-x} \right) 
{\bar{\epsilon}}_{2\nu} \left( 2E \delta^{\nu}_{\,\,0} + q^{\nu} \frac{x}{1-x} \right)   \nonumber \\
&& +  \frac{1}{x} ({\bar{\epsilon}}_1  {\bar{\epsilon}}_2)  
 \epsilon_{1\mu}  \left( 2E \delta^{\mu}_{\,\,0} - q^{\mu} \frac{x}{1-x} \right)  
 \epsilon_{2\nu}   \left( 2E \delta^{\nu}_{\,\,0} + q^{\nu} \frac{x}{1-x} \right) 
 \nonumber \\ \nonumber
 &&   \left. - \frac{1-x}{x^2}   (\epsilon_2 R {\bar{\epsilon}}_2)   \epsilon_{1\mu} 
 \left( 2E \delta^{\mu}_{\,\,0} - q^{\mu} \frac{x}{1-x} \right)
{\bar{\epsilon}}_{1\nu} \left( 2E \delta^{\nu}_{\,\,0} - q^{\nu} \frac{x}{1-x} \right)
  \right]~.
\end{eqnarray}
Finally the term with four momenta in the polarization contractions is given by
\begin{eqnarray}
\label{al2bis}
I_3 &=& \frac{(\alpha')^2}{4}\int_0^1 dx \,\,x^{-\alpha' s} (1-x)^{-\frac{\alpha'}{4} t } \left( \frac{1-x}{x} \right)^2 \epsilon_{1\mu} \left( 2E \delta^{\mu}_{\,\,0} - q^{\mu} \frac{x}{1-x} \right) \\ \nonumber 
&& {\bar{\epsilon}}_{1\nu}  \left( 2E \delta^{\nu}_{\,\,0} - q^{\nu} \frac{x}{1-x} \right)
\epsilon_{2\rho} \left( 2E \delta^{\rho}_{\,\,0} + q^{\rho} \frac{x}{1-x} \right) {\bar{\epsilon}}_{2 \sigma} 
\left( 2E \delta^{\sigma}_{\,\,0} + q^{\sigma} \frac{x}{1-x} \right)~ \ .
\end{eqnarray}
In order to derive these results, one needs to use momentum conservation along the longitudinal directions of the D-branes 
\begin{eqnarray}
(1+R) (p_1 +p_2)=0 \ , 
\label{momcon}
\end{eqnarray}
and the transversality condition ($\epsilon_1 p_1=\epsilon_2 p_2=0$) to rewrite scalar products such as $\epsilon_1 R p_2$ in terms of $q=p_1+p_2$ and $(p_1 + R p_1)^\mu = 2 E \delta^\mu_{~0}$. By assuming that the polarizations of the gravitons have a vanishing time component, we recover the results of~\cite{Klebanov:1995ni}. This simplification is automatic in the Regge limit: the presence of the combination $x/(1-x)$ in some terms of the integrand yields extra factors of $\alpha' s$ after the integration over $x$ is performed. Then in the combination $2E \delta^\mu_{~0} \pm q^\mu x/(1-x)$ we can neglect the first term (that scales like $E$) with the respect of the second (that scales like 
$E^2$). Thus in the Regge limit we have
\begin{eqnarray}
\label{Iz}
I_1 &\sim & (-\alpha' s)^{ 1 + \frac{ \alpha' t }{4} } \Gamma \left(- 1- \frac{\alpha' t }{4} \right) 
 (\epsilon_1 \epsilon_2)   ({\bar{\epsilon}}_1 {\bar{\epsilon_2}}) ~,
\\ \nonumber
I_2 &\sim & - \frac{\alpha'}{2} (-\alpha' s)^{ 1 + \frac{ \alpha' t }{4} }  \Gamma \left(- 1- \frac{\alpha' t }{4} \right)  \left[  (\epsilon_1 \epsilon_2)  ({\bar{\epsilon}}_1 q) ({\bar{\epsilon}}_2 q) +
({\bar{\epsilon}}_1 {\bar{\epsilon}}_2)  ({{\epsilon}}_1 q) ({{\epsilon}}_2 q) \right]~,
\\ \nonumber
I_3 &\sim & \frac{(\alpha')^2}{4} (-\alpha' s)^{ 1 + \frac{ \alpha' t }{4} }  \Gamma \left(- 1- \frac{\alpha' t }{4} \right)  (\epsilon_1 q) ({\bar{\epsilon}}_1 q) (\epsilon_2 q)  ({\bar{\epsilon}}_2 q) ~.
\end{eqnarray}
In conclusion we obtain~\eqref{calA1'}.

\section{Operator-eikonal exponentiation at annulus level}
\label{app:1loop}

In the closed string channel the annulus amplitudes can be evaluated
using two boundary states. For instance the one-loop correction to the amplitude in~\eqref{eq:diam} is
\begin{equation}
  \label{eq:diam1l}
  {\cal A}^{TT}_2 = C_{S^2} \frac{1}{4\pi} \kappa_{26}^2 \frac{\alpha'}{8\pi} N^2 \int d^2z_1 d^2z_2 \langle B |V_T(z_1) V_T(z_2) P |B \rangle~,
\end{equation}
where $P$ is the closed string propagator\footnote{In our conventions $d^2z = 2 d{\rm Re}z d{\rm Im}z$; this is the origin of the factor of $4$ on the second line of~\eqref{eq:diam1l2}.}
\begin{equation}
  \label{eq:clP}
  P = \frac{\alpha'}{2} (L_0 + \tilde{L}_0 -2)^{-1} = \frac{\alpha'}{8\pi} \int 
\frac{d^2\qr}{|\qr|^2}  
\qr^{L_0-1} \bar{\qr}^{\tilde{L}_0-1}~.
\end{equation}
The powers of $\kappa_{26}$ and $N$ count as usual the number of insertions and that of the borders, while the extra factor of $1/(4 \pi)$ follows from the residual symmetries of the annulus that, as in~\cite{D'Appollonio:2010ae}, we decided not to fix.

The contribution of the zero modes $q^\mu$ and $p^\mu$ to~\eqref{eq:diam1l} is
\begin{equation}
\label{Zerom}
(2 \pi)^{p+1} \delta^{(p+1)} (p_1 + p_2) (2 \pi^2 \alpha' \lambda)^{-(26-p-1)/2} {\rm e}^{\frac{\alpha'}{2 \pi\lambda} \left[
\left(s+ \frac{4}{\alpha'}\right)   \left(\log \frac{|z_1|}{|z_2|}\right)^2 - t \log |z_1| \log |z_2|  \right]} ~,
\end{equation}
where $\log|\qr|=-\pi \lambda$. Again we used~\eqref{eq:pqst} with  $m_1^2=m_2^2=-4/\alpha'$. The contribution of the non-zero modes has two effects. First it yields the usual annulus measure
\begin{equation}
  \label{eq:meas1l}
  d\mu_1 = 2\pi d\lambda \frac{1}{ |\qr|^2} \prod_{n=1}^{\infty} \frac{1}{\left(1 - |\qr|^{2n}\right)^{24}}~.
\end{equation}
Then it transforms the disk Green function $\log(z_i-z_j)$ into the annulus one, which can be
expressed in terms of the prime form $\log E(z_i,z_j)$, where
\begin{equation}
E(z_i, z_j ) =   (z_i - z_j)  \prod_{n=1}^{\infty} 
\frac{  \left( 1 - |\qr|^{2n} \frac{z_{i}}{z_{j}} \right) 
\left(1 - |\qr|^{2n} \frac{z_j}{z_i}  \right) }{( 1 - |\qr|^{2n})^2 }    \ , 
\label{Ez1z2}
\end{equation}
or in terms of the standard Jacobi $\theta$-function
\begin{equation}
E(z_i , z_j) = 2 \pi \ii  \,\,{\rm e}^{\ii \pi (\nu_i + \nu_j)} 
\frac{\theta_{1} ( \nu_i - \nu_j | \ii\lambda )  }{\theta_{1} ' (0 | \ii\lambda) }
\label{Eab}~.
\end{equation}
With these modifications, the evaluation of the annulus amplitude proceeds in a similar way to the disk amplitude in~\eqref{eq:TTd1}.
In order to follow closely the superstring derivation of~\cite{D'Appollonio:2010ae} we introduce the variables
\begin{equation}
z_{i } = {\rm e}^{2 \pi \ii \nu_{i}}\equiv {\rm e}^{2 \pi \ii (\ii \lambda \rho_1 - \omega_1)}  ~,~~~z_{j} = {\rm e}^{2 \pi \ii \nu_j}\equiv {\rm e}^{2 \pi \ii (\ii \lambda \rho_2 - \omega_2)} ~.
\label{zzk}
\end{equation}
The result is
\begin{eqnarray}
  \label{eq:diam1l2}
  {\cal A}^{TT}_2  & = &  \left(\frac{\kappa_{26} T_p N}{2}\right)^2 \frac{\alpha'}{16\pi}  (2 \pi^2 \alpha')^{-(25-p)/2} (2\pi)^4 
\int_{0}^{\infty}   \frac{d \lambda}{\lambda^{\frac{21-p}{2}}} \frac{1}{ |\qr|^2} \prod_{n=1}^{\infty} \frac{1}{\left(1 - |\qr|^{2n}\right)^{24}}
\\ \nonumber  & & 
 4 \int_0^{\frac{1}{2}} \!\! d \rho_1 \! \int_0^{\frac{1}{2}} \!\! d \rho_2 \! \int_0^1 \!\! d\omega_1 \! \int_0^1 \!\! d\omega_2  \,\, {\rm e}^{- (\alpha' s +2) V_s - \frac{\alpha' t}{4} V_t } \frac{ (\theta_1 ' (0 | \ii \lambda))^4\,\, {\rm e}^{ 4 \pi \lambda \rho^2}  }{(2\pi)^4 \theta_1^2 ( \ii \lambda \rho - \omega |  \ii \lambda )  \theta_1^2 (- \ii \lambda \rho - \omega | \ii \lambda)}~,
\end{eqnarray}
where 
\begin{equation}
 \rho = \rho_1 - \rho_2 \ , \hspace{1cm} \omega = \omega_1 - \omega_2 \ , \hspace{1cm}\zeta= \rho_1 + \rho_2 \ , 
\label{rhoomegazeta}
\end{equation}
and
\begin{eqnarray}
V_s & = & - 2 \pi \lambda \rho^2 + \log  \frac{ \theta_1 ( \ii \lambda (\zeta+\rho) | \ii \lambda) \theta (\ii \lambda (\zeta- \rho)| \ii \lambda)}{ \theta_1 ( \ii \lambda \zeta+\omega | \ii \lambda) \theta (\ii \lambda \zeta- \omega| \ii \lambda)}~,    \nonumber \\
V_t & = & 8 \pi \lambda \rho_1 \rho_2 + \log  \frac{ \theta_1 ( \ii \lambda \rho + \omega | \ii \lambda) \theta (\ii \lambda  \rho - \omega | \ii \lambda)}{ \theta_1 ( \ii \lambda \zeta+\omega | \ii \lambda) \theta (\ii \lambda \zeta- \omega| \ii \lambda)}~.
\label{VsVtVm}
\end{eqnarray}
As in the superstring case, in our kinematic configuration (large $E$ and small $R_p/b$), this integrand is dominated by the 
region of small $\rho$ and large $\lambda$. In this limit we have
\begin{eqnarray}
 {\cal A}^{TT}_2 &\sim &  \left(\frac{\kappa_{26} T_p N}{2}\right)^2 \frac{\alpha'}{16\pi}  (2 \pi^2 \alpha')^{-(25-p)/2} (2\pi)^4
  \int_{0}^{\infty}   \frac{d \lambda}{\lambda^{\frac{21-p}{2}}} \,\,\, {\rm e}^{2   \pi \lambda }  
  \int_0^1 d \zeta \int_0^1 d \omega  \nonumber \\
&& 2 \int_{-\zeta}^{\zeta} d \rho \,\, {\rm e}^{2 \pi \alpha' \lambda s \rho^2}
{\rm e}^{ 2 \pi \lambda \zeta (1 - \zeta) \frac{\alpha' t}{4} } ( 4 \sin^2 \pi \omega)^{- \frac{\alpha' t}{4}} 
\nonumber \\
&& 
\exp \left[4 \alpha' s \sin^2 (\pi \omega) \left( {\rm e}^{- 2 \pi \lambda \zeta} + {\rm e}^{- 2 \pi \lambda(1-  \zeta)} \right) \right]  \left(4 \sin^2 (\pi \omega)\right)^{-2}~.
\label{rho=0}
\end{eqnarray}
The last term in the last line comes from the last fraction in Eq.~(\ref{eq:diam1l2}). The integral over $\rho\sim 0$ is Gaussian (after a Wick rotation $E\to \ii E_e$). By writing the exponential in the last line as a double series of terms proportional to $e^{-2\pi n \lambda \zeta}$ and $e^{-2\pi m \lambda (1-\zeta)}$ we obtain an expression very similar to the integrand $I_1$ in Appendix~A of~\cite{D'Appollonio:2010ae}. Then the integral over $\omega$ can also be performed and one obtains
\begin{eqnarray}
{\cal A}^{TT}_2 & \sim & \left(\frac{\kappa_{26} T_p N}{2}\right)^2 \frac{\alpha'}{8\pi}  (2 \pi^2 \alpha')^{-(25-p)/2} (2\pi)^4
  \int_0^1 d \zeta     \frac{\ii}{\sqrt{2 \alpha' s}}  (\alpha' s)^{2} 4^{- \frac{\alpha' t}{4}}    \nonumber \\
& &    \sum_{n,m=0}^{\infty} \frac{1}{n! m!}  \left( 4 \alpha' s \right)^{(n-1) + (m-1) - \frac{\alpha't}{4}}   \frac{1}{\pi} B\left(\frac{1}{2} , n+m-2  - \frac{\alpha't}{4} + \frac{1}{2} \right) \label{gty4}
 \\
& &  \Gamma\left(  \frac{p}{2} -10  \right)  \left[ -2 \pi  \zeta (1- \zeta) \frac{\alpha' t}{4} +  2\pi(n-1) \zeta
 + 2 \pi (m-1)  (1- \zeta) \right]^{10 - p/2} \ . \nonumber
\end{eqnarray}
This expression is very similar to the superstring case, 
except that $n,~m$ are shifted to $n-1,~m-1$, due 
to the presence of the 
tachyon pole ($e^{2\pi\lambda}$) in the first line of~\eqref{rho=0} and of the last factor ($sin^{-4} (\pi \omega)$) in the final line of the same equation. We can trade the integral over $\zeta$ for a momentum integral in $D=24-p$ dimensional space by using the identity
 \begin{eqnarray}
\int_0^1 d \zeta \,\, \Gamma ( \frac{p -20}{2} )   \left[ -2   \zeta (1- \zeta) \frac{\alpha' t}{4} +  2 (n-1) \zeta + 2  (m-1)  (1- \zeta) \right]^{\frac{20 - p}{2}} \nonumber \\
 = (2 \pi \alpha' )^{\frac{24- p}{2}}  \int  \frac{ d^{24-p} \mathbf{k}}{ (2\pi)^{24-p}} 
\left[ 2(n-1) + \frac{\alpha'}{2} \mathbf{k}^2\right]^{-1} \left[ 2(m-1) + \frac{\alpha'}{2} (\mathbf{k}-\mathbf{q})^2\right]^{-1} ~.
 \label{polesr}
\end{eqnarray}
We can rewrite the sum as an integral 
\begin{equation}
  \label{eq:sommer}
  \sum_{m=0}^\infty \frac{1}{m!} \frac{f(m) s^m}{m+t} = 
  - \int_{\cal C}\frac{dm}{2\pi \ii} {\rm e}^{-\ii \pi m} \Gamma(-m)  \frac{f(m) s^m}{m+t}~,
\end{equation}
where the contour includes all the poles in the $\Gamma(-m)$ and not the other ones. We can then
deform the contour and focus on the poles of the propagators in~\eqref{polesr}, which are
the only ones that contribute to the leading term in the energy. We find
\begin{eqnarray}
{\cal A}^{TT}_2 & \sim &
\left(\frac{\kappa_{26} T_p N}{2}\right)^2 \frac{\alpha'}{8\pi}  (2 \pi^2 \alpha')^{-(25-p)/2} (2\pi)^4 
      \frac{\ii \pi^{10- p/2}}{\sqrt{2 \alpha' s}} \frac{1}{4} (\alpha' s)^{2} 4^{- \frac{\alpha' t}{4}}    \nonumber \\
&& (2 \pi \alpha' )^{\frac{24 - p}{2}}  \int  \frac{ d^{24-p} \mathbf{k}}{ (2\pi)^{24- p}}   
\Gamma \left( -1 + \frac{\alpha'}{4} \mathbf{k}^2\right ) \Gamma \left( -1 + \frac{\alpha'}{4} (\mathbf{q}-\mathbf{k})^2 \right) {\rm e}^{\ii \pi \frac{\alpha'}{4} \mathbf{k}^2 + \ii \pi \frac{\alpha'}{4} (\mathbf{q}-\mathbf{k})^2}  \nonumber \\
&& \left( 4 \alpha' s \right)^{- \frac{\alpha'}{4} \mathbf{k}^2 -   \frac{\alpha'}{4}{(\mathbf{q}-\mathbf{k})^2}}   \frac{1}{\pi} B\left(\frac{1}{2} , - \frac{\alpha'}{4} \mathbf{k}^2  -  \frac{\alpha'}{4} (\mathbf{q}-\mathbf{k})^2   - \frac{\alpha't}{4} + \frac{1}{2}\right) ~.
\label{fibuy}
\end{eqnarray}
By using Eq.~\eqref{Atach} and
\begin{eqnarray}
\frac{1}{\pi}B \left( \frac{1}{2} + \frac{\alpha'}{4} \mathbf{k}(\mathbf{q}-\mathbf{k}) ,  \frac{1}{2} \right) = 2^{- \alpha' \mathbf{k} (\mathbf{q}-\mathbf{k})} \langle 0| \prod_{i=1}^{2} \int\limits_0^{2 \pi} \frac{ d \sigma_i}{2 \pi} :{\rm e}^{\ii \mathbf{k} {\hat{X}}(\sigma_1)} :\,:{\rm e}^{\ii (\mathbf{q}-\mathbf{k}) {\hat{X}}(\sigma_2)} : | 0 \rangle~,
\label{BBBB}
\end{eqnarray}
we can rewrite~\eqref{fibuy} as follows
\begin{eqnarray}
{\cal A}^{TT}_2 & \sim &  \frac{\alpha'}{8\pi}  (2 \pi^2 \alpha')^{-(25-p)/2} (2\pi)^4 \frac{\ii \pi^{10- p/2}}{\sqrt{2 \alpha' s}} \frac{1}{4} (2 \pi \alpha' )^{\frac{24 - p}{2}}
\int  \frac{ d^{24-p} \mathbf{k} }{ (2\pi)^{24- p}} 
\\ \nonumber &&
A_1(s,\mathbf{k}) A_1(s,\mathbf{q}-\mathbf{k}) \langle 0| \prod_{i=1}^{2} \int\limits_0^{2 \pi} \frac{ d \sigma_i}{2 \pi} :{\rm e}^{\ii \mathbf{k} {\hat{X}}(\sigma_1)} :\,:{\rm e}^{\ii (\mathbf{q}-\mathbf{k}) {\hat{X}}(\sigma_2)} : | 0 \rangle~.
\end{eqnarray}
Since the normalization on the first line is just $\ii/(4 E)$, we indeed obtain~\eqref{eq:conv} for $h=2$. 

\section{Shapiro time delay}
\label{app:Shapiro}

In this Appendix we compute the Shapiro time delay for  a probe particle moving in the metric created by a stack of $N$ D$p$-branes. In order to include also the case of the D$p$-branes of the bosonic string we write the metric~\footnote{See Refs.~\cite{Duff:1994an,Argurio:1998cp} for a derivation of this metric. See also Ref.~\cite{DiVecchia:1997pr}.}  keeping an arbitrary space-time dimension $d$
\begin{eqnarray}
ds^2 = ( H(r))^{- \frac{d-p-3}{d-2}} \left( - dt^2 + dx_p^2 \right) + ( H (r))^{\frac{p+1}{d-2}} dx_{d-1-p}^2 \ ,
\label{ds2d}
\end{eqnarray}
where
\begin{eqnarray}
H(r) = 1 + \left( \frac{R_p}{r} \right)^{d-p-3} \ , \hspace{1cm} R^{d-p-3}_p = \frac{2 \kappa_d T_p N}{(d-3-p) \Omega_{d-p-2}} \ .
\label{Hr}
\end{eqnarray}
$T_p$ is given in Eq. (\ref{eq:Tp}) and $\kappa_d$ is the gravitational constant in $d$ dimensions
\begin{eqnarray}
2 \kappa_{d}^{2} = \frac{1}{2^{\frac{d-10}{2}}}  g_s^2 (2\pi)^{d-3} (\alpha')^{\frac{d-2}{2}}~~~.
\label{kappa2}
\end{eqnarray}
For $d=10$ the metric reduces to that of the maximally supersymmetric D$p$-branes in ten dimensions. Using Eqs.  (\ref{eq:Tp}) and (\ref{kappa2}), we can write the combination appearing in the  amplitudes in Sect. \ref{2} in terms of $R_p$:
\begin{eqnarray}
  \frac{1}{2} \kappa_d T_p  N  =  \frac{  \pi^{\frac{d-p-1}{2}}  R_{p}^{d-p-3}}{ \Gamma ( \frac{d-p-3}{2})  } \ , \hspace{1cm}
R_p^{d-p-3} =  \frac{ \Gamma ( \frac{d-p-3}{2}) }{ \pi^{\frac{d-p-1}{2}}  } \frac{g_s N}{4}  \frac{1}{2^{\frac{d-10}{2}}} (2 \pi \sqrt{\alpha'} )^{d-p -3} \,\, .
\label{Rp}
\end{eqnarray}
Let us assume that the probe particle moves along one of the transverse directions $x^{d-1} \equiv z$. In general, moving in the metric of the branes, its trajectory  will be deflected. However, if the impact parameter is much larger than $R_p$
the deflection angle will be very small and, in the first approximation, we can neglect it and assume that the probe moves along $z$. In this case we can also expand the metric for large $r$ and keep only the dominant terms. We then obtain the following relation between $dt$ and $dz$
\be
- dt^2 \left( 1 + \frac{d-p-3}{d-2}  \left( \frac{R_p}{r} \right)^{d-p-3} + \dots \right)
 + dz^2 \left(1 + \frac{p+1}{d-2}   \left( \frac{R_p}{r} \right)^{d-p-3} + \dots \right) =0 \ , \label{ds2b}
\ee
which  implies
\begin{eqnarray}
\frac{dt}{dz} =  1 + \frac{1}{2} \left( \frac{R_p}{r} \right)^{d-p-3}+ \dots \ , 
\label{dtdz}
\end{eqnarray}
where $r^2 =b^2 +z^2$.  From this expression we can immediately compute the Shapiro time delay
\begin{eqnarray}
\Delta t =  \frac{   R_p^{d-p-3} }{2} 
\int_{-\infty}^{+\infty} \frac{dz}{ (b^2 + z^2)^{\frac{d-p-3}{2} }} = \frac{ R_p^{d-p-3} \sqrt{\pi} \Gamma ( \frac{d-p-4}{2})}{ 2 
\mathbf{b}^{d-p-4} \Gamma ( \frac{d-p-3}{2})} \ . 
\label{deltatbis}
\end{eqnarray} 
On the other hand, if we compute the quantity in Eq. (\ref{tach}) keeping only the pole of the graviton we get
\begin{eqnarray}
2\delta (s,b) &=& \int \frac{d^{d-2-p} \mathbf{q}}{(2\pi)^{d-2-p}} \frac{A_1 (s, \mathbf{q})}{2E}   {\rm e}^{\ii \mathbf{q} \mathbf{b}} = \frac{\pi^{\frac{d-p-1}{2}} R_p^{d-p-3} 2E}{\Gamma ( \frac{d-p-3}{2})} 
 \int \frac{d^{d-2-p} \mathbf{q}}{(2\pi)^{d-2-p}}  \frac{1}{\mathbf{q}^2} \nonumber \\
   &=& \frac{ R_p^{d-p-3} \sqrt{\pi}\Gamma ( \frac{d-p-4}{2})}{
  2 \mathbf{b}^{d-p-4} \Gamma ( \frac{d-p-3}{2}) } \, E \ , 
\label{deltavv}
\end{eqnarray}
which implies the following relation between the phase shift and the Shapiro time delay
\begin{eqnarray}
\Delta t = \frac{\partial}{\partial E}  2\delta (s, \mathbf{b}) \ . 
\label{deltatdelta}
\end{eqnarray}
In order to have a time delay without deflection, we could have studied, following~\cite{Camanho:2014apa}, the propagation
of a probe particle between two equidistant stacks of D$p$-branes,  displaced for instance from the origin only along the direction 
$x^{d-2}$ (i.e. located at $x^{d-2} = -b$ and at $x^{d-2} = b$).  
The result would have been twice the value of the time delay derived in this Appendix.

\end{appendix}

\providecommand{\href}[2]{#2}\begingroup\raggedright\endgroup

\end{document}